\documentclass[a4paper,11pt]{article}
\usepackage{graphicx}
\usepackage{typearea}
\title{Pre-seismic ionospheric anomalies detected before the 2016 Kumamoto earthquake}
\author{Takuya Iwata, Ken Umeno}
\date{}
\typearea{16}
\begin{document}
\maketitle
Department of Applied Mathematics and Physics,
Graduate School of Informatics, Kyoto University, Kyoto, Japan

%----------------------------------------------------------------------------------------
%	ABSTRACT
%----------------------------------------------------------------------------------------

% Do NOT include any \begin...\end commands within the body of the abstract.

\begin{abstract}

On April 15, 2016, the Kumamoto earthquake (Mw 7.3) occurred in Japan with no warning signals.
Global Navigation Satellite System (GNSS) receivers provide useful information on disturbances in ionosphere by calculating the changes in Total Electron Content (TEC), which is the number of electrons in ionosphere.
Here we show our recently proposed correlation analysis of TEC data which can detect the pre-seismic ionospheric anomalies from the public GNSS data.
Our method detected the ionospheric anomaly several tens of minutes before the 2016 Kumamoto earthquake near its epicenter.
Furthermore, we gave an indicator to distinguish between the pre-seismic TEC anomalies and the medium scale traveling ionospheric disturbances (MSTIDs) by calculating the anomalous area rates.
These results support the hypothesis for existence of the preceding phenomena before large earthquakes.
\end{abstract}

%----------------------------------------------------------------------------------------
%	ARTICLE CONTENT
%----------------------------------------------------------------------------------------

% The body of the article must start with a \begin{article} command
% \end{article} must follow the references section, before the figures and tables.

\section{Introduction}
Ionosphere, a shell of a large amount of electrons spreading above us, is disturbed by various natural phenomena such as volcanic eruptions \cite{Heki2006, Dautermann2009}, solar flares \cite{Donnelly1976}, earthquakes \cite{Calais1995, Astafyeva2011, Cahyadi2015}, and so on.
Observing the ionospheric disturbances provides miscellaneous information on the condition of the upper atmosphere \cite{Ho1996}.
Japan has a dense GNSS observation network \cite{Saito1998} (GNSS earth observation network, GEONET) and the GNSS data from this network are available freely via the Internet (terras.gsi.go.jp).
Daily TEC values are estimated by calculating the phase differences between the two carrier waves from the GNSS satellites \cite{Sardon1997, Mannucci1998}.
TEC data are useful for monitoring the conditions of the ionosphere, and are used by many researchers.
Hypotheses for preceding phenomena of earthquakes include the ionospheric disturbances before large earthquakes \cite{Heki2011, Heki2015}.
As the hypothesis is still under discussion, we have to verify whether the hypothesis is right or not \cite{Kamogawa2011, Utada2011, Heki2013, Masci2015}.
We can get typically two TEC time series from a pair of a satellite and a GNSS station on the ground because each GPS satellite is seen twice per day at each GPS receiver.
Since \verb+~+30 satellites are available in orbit, we can get over 60,000 time series in a day from GEONET, composed of about 1300 GNSS stations.

%------------------------------------------------

\section{Methods}

% \subsection{Convert RINEX data to TEC data}

% Observation data from GEONET are derived as RINEX data21.  We have to convert RINEX data to TEC data to analyze.  There are many researches using TEC data and the methods are open to the public.  In this research, we used the FORTRAN programs developed by Professor Heki at Hokkaido University.  This source code is open at his HP (http://www.ep.sci.hokudai.ac.jp/~heki/software.htm).

%\subsection{Correlation analysis \cite{Iwata2016}}

First of all, we choose a GNSS station as a “central station”.
Next, we calculate the distances between the central station and the other GNSS stations in Japan to pick up the 30 stations which are the nearest to the central station and call the 30 stations “surrounding stations”.
We number the central station and each surrounding stations from 0 to 30, where the number 0 means the central station and the numbers 1 to 30 are allotted to the surrounding stations.
Since the data length of the TEC data in a day varies by station, we extract the data from the TEC data at the center station and the surrounding stations that have common time information.
We set up parameters $t_{sample}$ and $t_{test}$.  In this research, we set up $t_{sample} = 2.0$ [hours] and $t_{test} = 0.25$ [hours].
At each station $i$ and each time epoch $t$, let Sample Data be the TEC data from time $t$ to $t + t_{sample}$ and Test Data be the TEC data from time $t + t_{sample}$ to $t + t_{sample} + t_{test}$.
That is, the time length of the Sample Data is $t_{sample}$ and the time length of the Test Data is $t_{test}$.
At each station, we fit a reference curve to Sample Data by the least square method.
We can choose some kinds of reference curves here, but the result does not highly depend on the choice of reference curve \cite{Iwata2016}.
In this research, we choose the 7th polynomial functions as the reference curves.
In this way, we get the reference curves from the Sample Data.
Next, we calculate a deviation of the Test Data from the reference curve at each station.
The deviation at station $i$ at time $t$ is denoted as $x_{i, t}$.
This value $x_{i, t}$ means the abnormality of TEC at each station.
Finally, we calculate a summation of correlations $C(T)$ between the deviations at the central station and the surrounding stations as follows:
\begin{equation}
C(T) = 1/(N * 30) \sum_{i=1}^{30} \sum_{j=0}^{N-1} x_{i,t + t_{sample}  + j\Delta t} x_{0,t + t_{sample}  + j\Delta t}
\end{equation}
\[
T = t + t_{sample} + t_{test}
\]
Here, $N$ is the number of data in Test Data, $\Delta t$ is a sampling interval in Test Data, which means $\Delta t =  t_{test}/(N-1)$.
Note that we do not use the TEC data after time epoch $T$ to calculate $C(T)$.

%------------------------------------------------

\section{Results}

\subsection{Correlation analysis for the 2016 Kumamoto earthquake}

\subsubsection{Correlation values all over Japan on 15th April, 2016}

\begin{figure}
\includegraphics[width=0.8\linewidth]{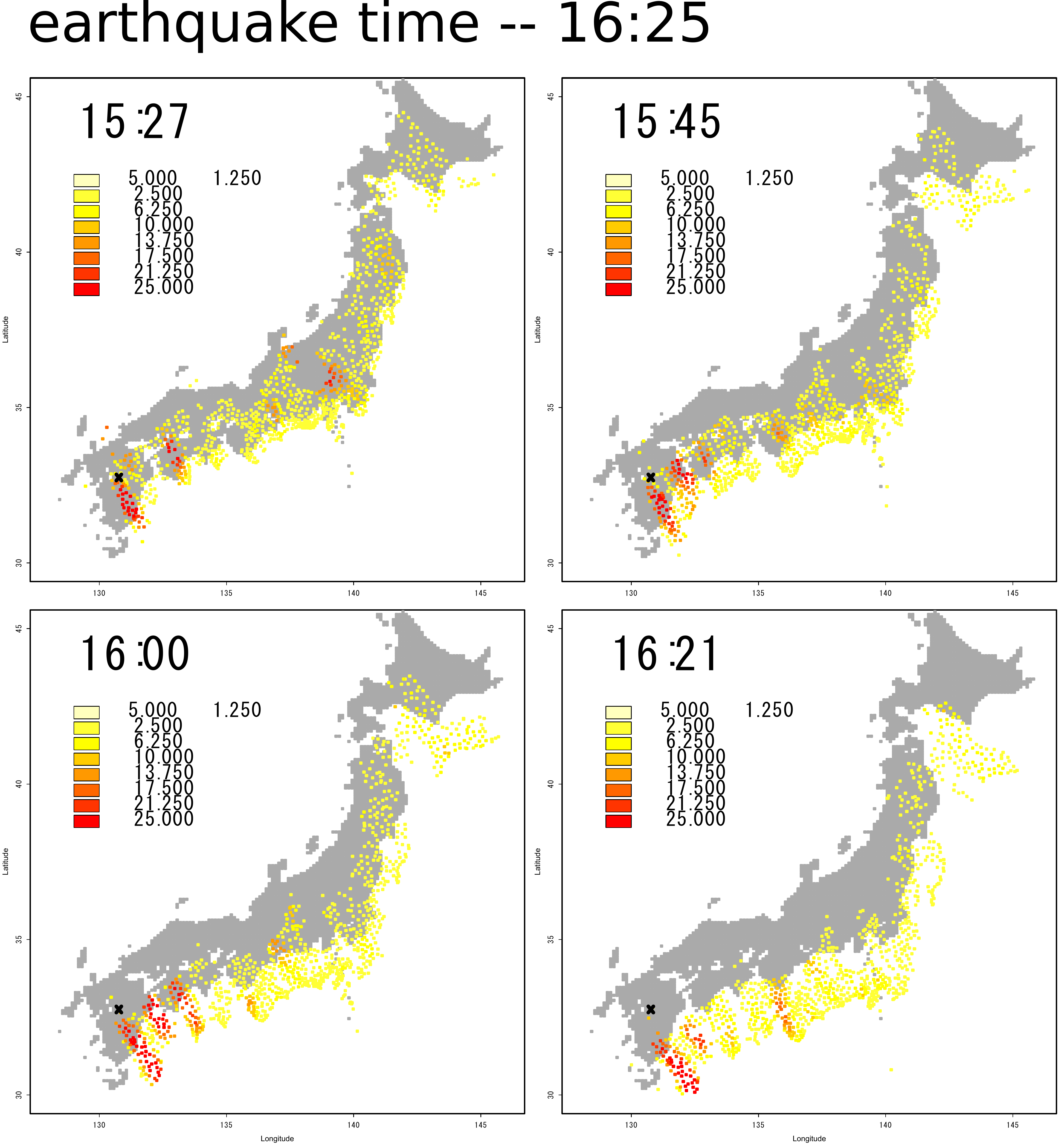}
\caption{Correlation values at all GNSS stations in Japan before the 2016 Kumamoto earthquake.  We used every GNSS station as a central station and mapped the results into the Japan map.  The GPS satellite 17 is used here.  Red points in the figures show high correlation values (anomalous).  The black x marks represent the epicenter of the earthquake.  The earthquake occurrence time is 16:25 [UT] and each time in the figures corresponds to about one hour, 40 minutes, 25 minutes and 5 minutes before the main shock.}
\label{correlation}
\end{figure}

The 2016 April 15 (16:25 UT) Kumamoto earthquake occurred in southwest Japan.
Its magnitude is Mw 7.3 and this earthquake severely damaged to around Kumamoto Prefecture.
We analyzed the TEC data on the day and found that our correlation analysis captured the TEC anomalies near the epicenter about one hour before the main shock (Fig. \ref{correlation}).
Our correlation analysis is an analysis method to calculate the correlation of TEC anomalies between one GNSS station and the other GNSS stations surrounding it \cite{Iwata2016}.
This method is effective because the noises on the data can be effectively suppressed by the statistical superposition effect and the genuine anomalies are emphasized by calculating the correlation.
Here, we used GPS (Global Positioning System, American GNSS) satellite 17 and all the GNSS stations in Japan.
The elevation mask of the line-of-sight is $15^\circ$.
The red points in Fig. \ref{correlation} mean high TEC correlation values (anomalous) and the yellow points mean calm condition (normal).
The location of each point on this Japan map indicates the intersection of ionosphere with the line of sight between the satellite and each GNSS station (Sub Ionospheric Point, SIP).
For simplification, we make an assumption that ionosphere is a thin layer about 300 kilometers above us.
Therefore, the points in Fig. \ref{correlation} slightly move every minute as the satellite goes around the earth.
Since the ionosphere is subject to the natural phenomena, the average TEC values in Japan are continuously changing as a function of the time.
For example, it is known that enhanced ultraviolet flux by solar flares causes sudden increase of electron densities in ionosphere in a wide area \cite{Leonovich2002}.

\subsubsection{Correlation values near the epicenter}
\begin{figure}
\includegraphics[width=0.8\linewidth]{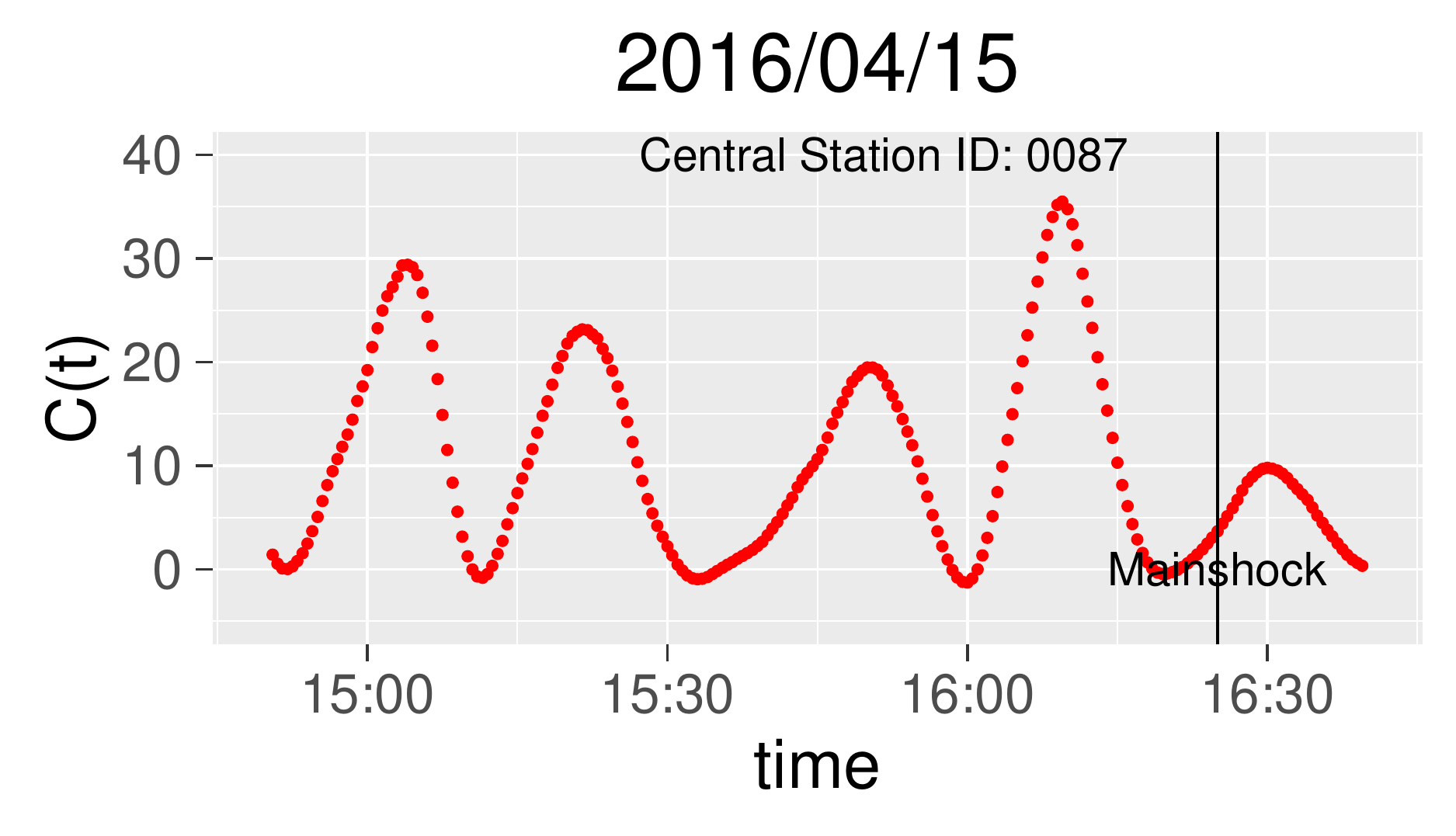}
\caption{Correlation values before the 2016 Kumamoto earthquake.  The vertical axis shows the correlation $C(T)$ and the horizontal one the time $t$ [UTC]. The black line indicates the exact time 16:25 [UTC] when the 2016 Kumamoto earthquake occured.  We used the pair of the 0087 (Koga, Fukuoka Prefecture) GNSS station as a central station and GPS satellite 17. }
\label{correlation_local}
\end{figure}

\begin{figure}
\includegraphics[width=0.8\linewidth]{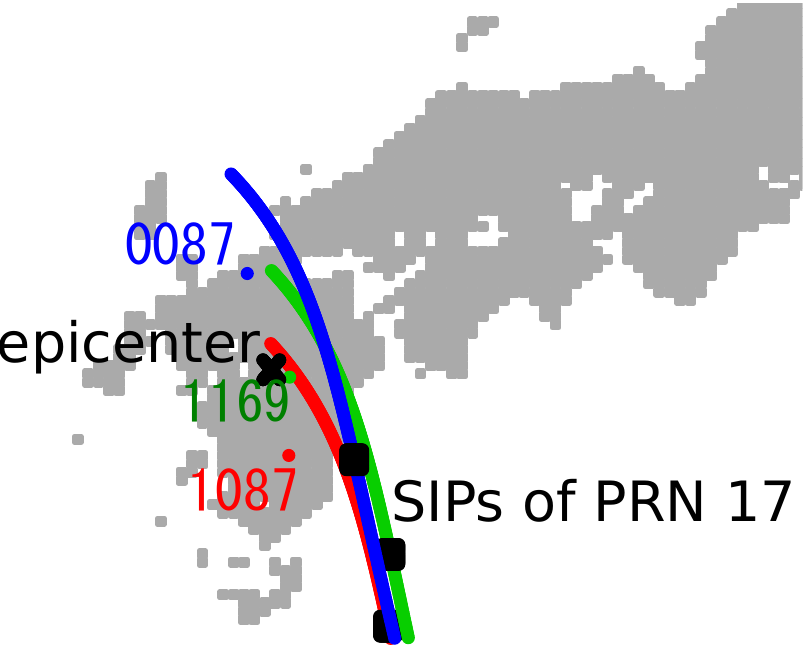}
\caption{The geographical position relations of a single GPS (PRN 17) satellite pass (UT 14:00-17:00) seen from a few representative receivers, including the central GNSS station (ID:0087) used in Fig. \ref{correlation_local}.  The colored lines represent the SIP tracks and the black points on them are the SIPs when the earthquake occured.}
\label{location}
\end{figure}

Figure \ref{correlation_local} shows the result of the correlation analysis on the earthquake day near the epicenter.
We chose the GPS satellite 17 and the 0087 (Koga, Fukuoka Prefecture) GNSS station as a central station.
The x-axis is universal time and the y-axis represents an accumulated correlation, here, briefly written $C(T)$ defined in Eq. (1).
The black line represents T=16:25 [UT], i.e. the time when the 2016 Kumamoto earthquake occured.
A typical example for a single GPS (PRN 17) satellite pass seen from a few representative receivers, including the central GNSS station (ID:0087) used in Fig. \ref{correlation_local}, is shown in Fig. \ref{location}.
The maximum distance between the 30 receivers and the central receiver (ID:0087) is about 78.8 km on the ground and the maximum distance of associated SIPs of PRN 17 is about 72.7 km.

\subsubsection{Correlation values on the days before the earthquake}
\begin{figure}
\includegraphics[width=0.8\linewidth]{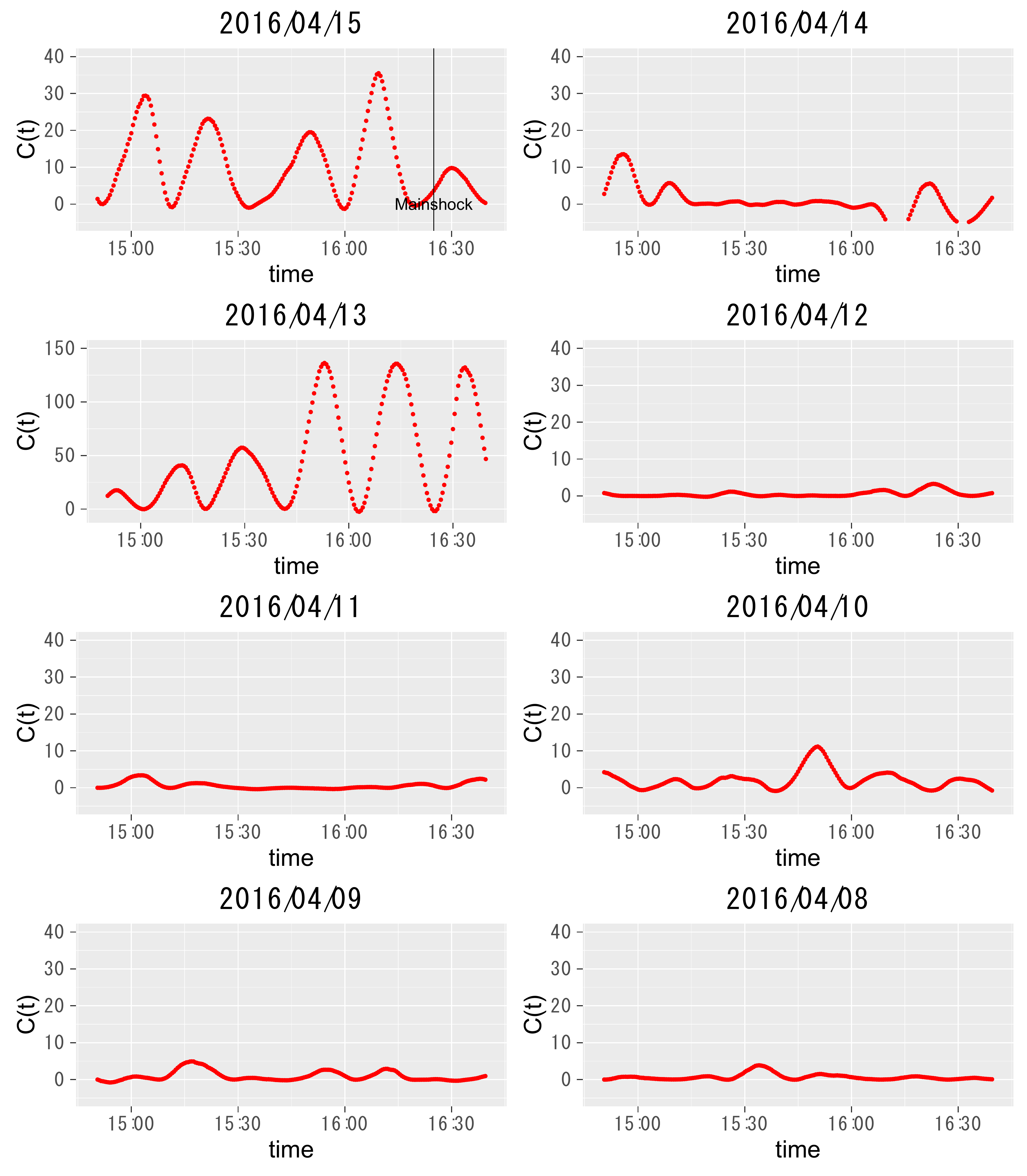}
\caption{Correlation values near the epicenter on the non-earthquake days and the earthquake day.  We used the pair of GPS satellite 17 and the GNSS station 0087 (Koga, Fukuoka Prefecture), which is near the epicenter when the earthquake occurred.  The x-axis shows time [UT] and the y-axis shows a correlation value at each time.  The black vertical line on 2016/04/15 represents the earthquake occurrence time 16:25 [UT].}
\label{non-earthquake}
\end{figure}

\begin{figure}
\includegraphics[width=0.8\linewidth]{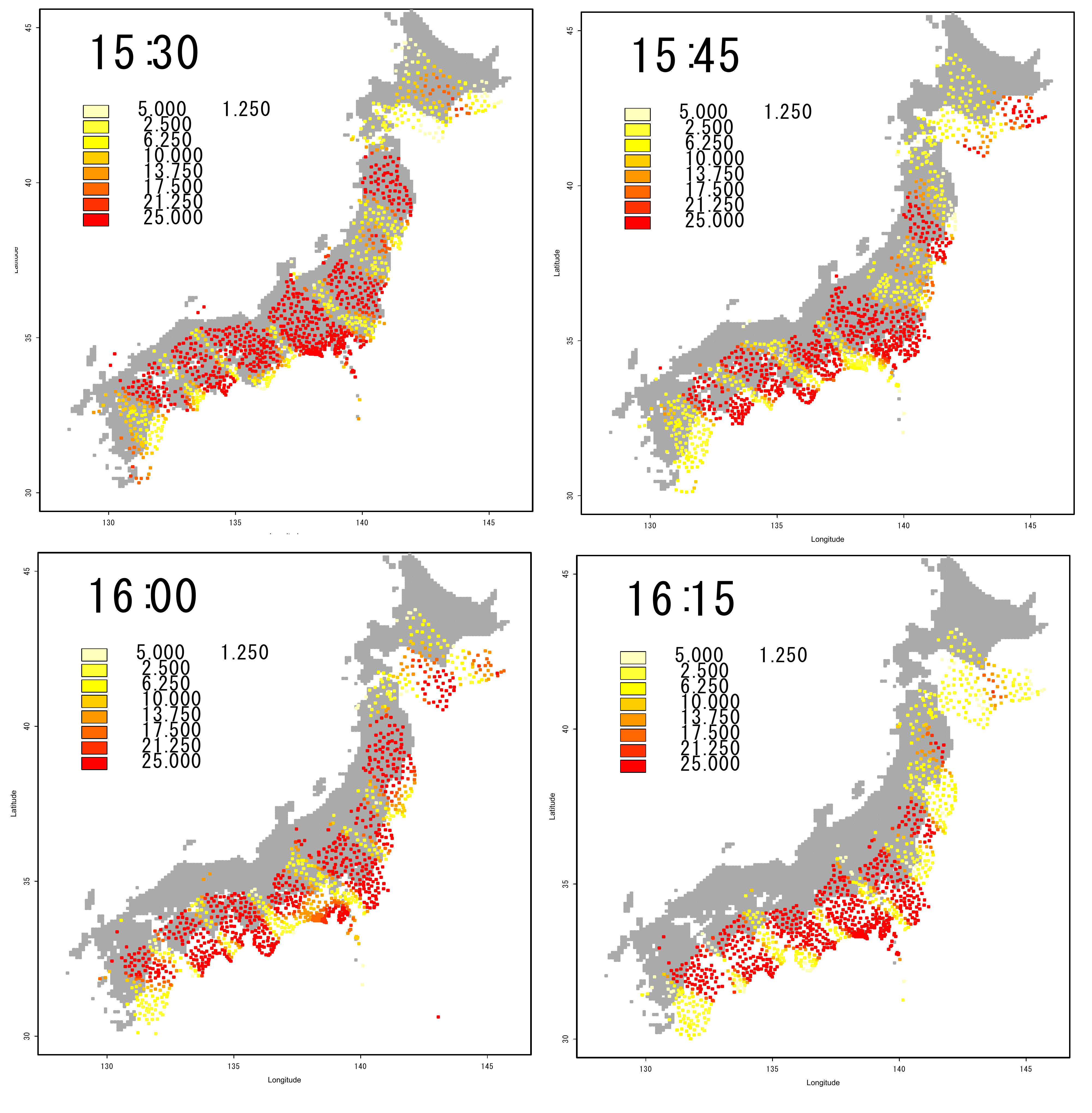}
\caption{Correlation values at all GNSS stations in Japan on 2016/04/13.  We used every GNSS station as a central station and mapped the results into the Japan map.  The GPS satellite 17 is used here.}
\label{mstid}
\end{figure}

Generally, when we think of anomality detection, it is important to confirm that how often the method makes a wrong detection in a normal condition (type I error, false positive) as well as the method does not make a right detection in an abnormal condition (type II error, false negative).
To confirm them, we calculated the TEC correlation on non-earthquake days as well (Fig. \ref{non-earthquake}).
This figure shows the correlation values at the GNSS station near the epicenter (ID:0087, Koga, Fukuoka Prefecture) with the satellite 17 from 2016/04/08 to 2016/04/15.
2016/04/15 is the day the Kumamoto earthquake occurred at 16:25 [UT].
In comparison to the correlation values on the earthquake day, the correlation values on the other days except 2016/04/13 are quite small and actually quiet.
The most correlation values on these days are less than 10, whereas the correlation values are more than 30 about 20 minutes before the main shock.
However, the correlation values on 2016/04/13 are excessively large in spite that no large earthquakes occurred on that day.
Such anomalous correlation values should be caused by the medium scale traveling ionospheric disturbances (MSTIDs).
Figure \ref{mstid} shows the result of correlation analysis on 2016/04/13.
We can see the ionospheric disturbances in a wide range on the day.
These TEC anomalies move from northeast to southwest at a speed of 100-200 m/s, which is consistent with the previous study of MSTIDs \cite{Otsuka2011}.
Distinguishing between the pre-seismic TEC anomalies and the MSTIDs is one major obstacle to establishing an earthquake prediction method from ionospheric conditions.
One of the methods to solve this problem is shown in the next subsection.
These abnormal correlation values also show the characteristic wave-like patterns, which can be seen also in the 2011 Tohoku-Oki earthquake \cite{Iwata2016}.

\subsection{Distinction of earthquake precursors and space weather signals}

\subsubsection{Seasonal MSTIDs and the 2016 Kumamoto earthquake}

The ionospheric disturbances before the Kumamoto earthquake look like the seasonal MSTIDs at first, which can be caused by atmospheric gravity waves that propagate upward from the lower atmosphere, or created in conjunction with auroral activity \cite{Otsuka2011}.
The statistical study of MSTIDs shows that the MSTIDs occurrence rate in Japan strongly depends on the season and local time \cite{Otsuka2011}.
According to their research, a high occurrence rate can be seen at four regions: dawn (05:00-07:00 LT) in summer (May-August), daytime (08:00-12:00 LT) in winter (November-February), dusk (17:00-20:00 LT) in summer, and nighttime (21:00-03:00 LT) in summer.
%As the 2016 Kumamoto earthquake occurred at nighttime in equinox (March, April, September and October), there is a relatively low possibility that the ionospheric disturbances before the earthquake is a mere MSTID and has few connections with the Kumamoto earthquake [Fig. 21.2 in \textit{Otuka et al.} (2011)].\\

\begin{figure}
\includegraphics[width=0.8\linewidth]{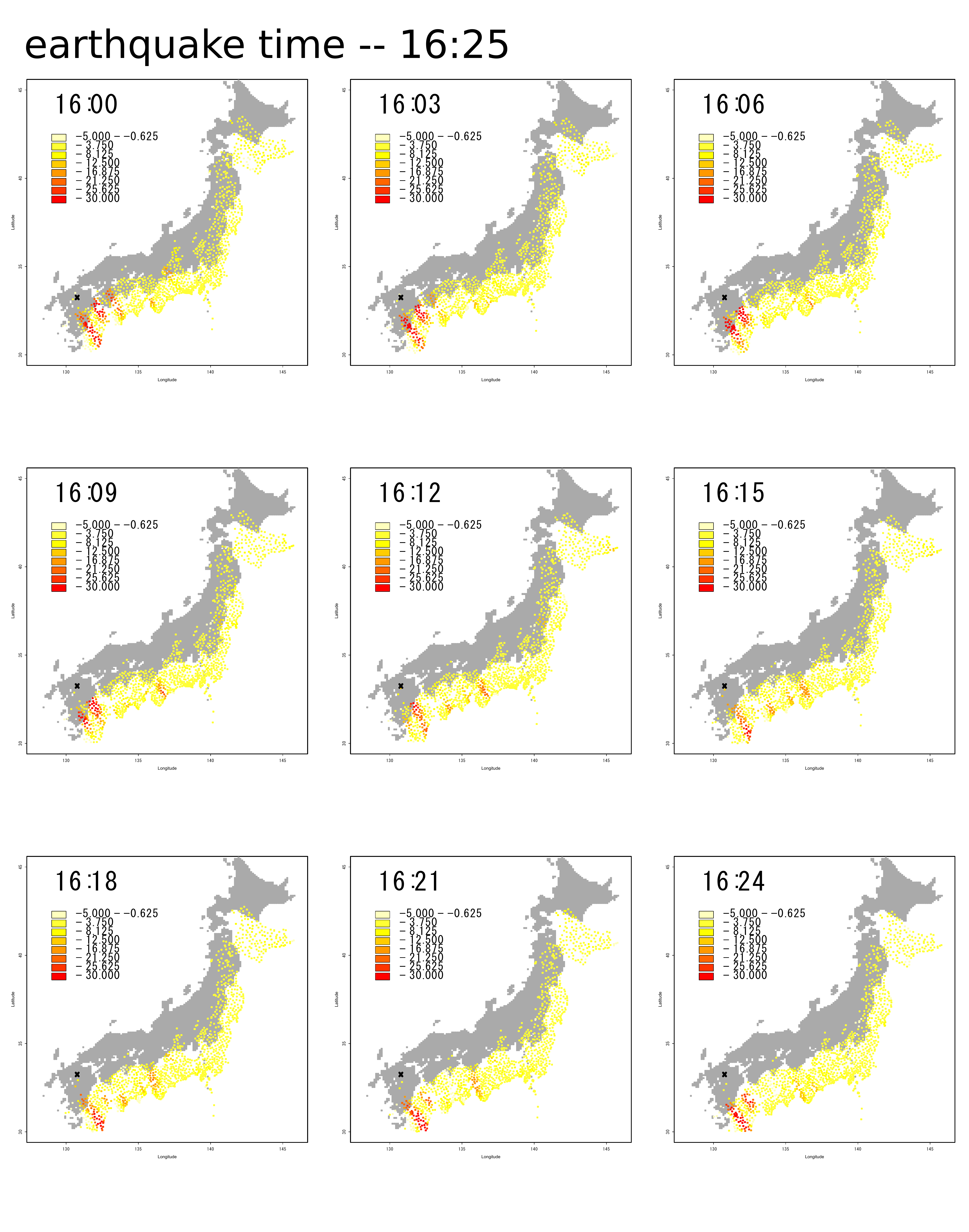}
\caption{Correlation values of PRN 17 over Japan with smaller time step (3 minutes) from 16:00UT to 16:24UT on 2016/04/15.  The earthquake occurrence time is 16:25 [UT].  The red area in each map means TEC anomalous area.  The x mark represents the epicenter.	}
\label{smallstep}
\end{figure}

\begin{figure}
\includegraphics[width=0.8\linewidth]{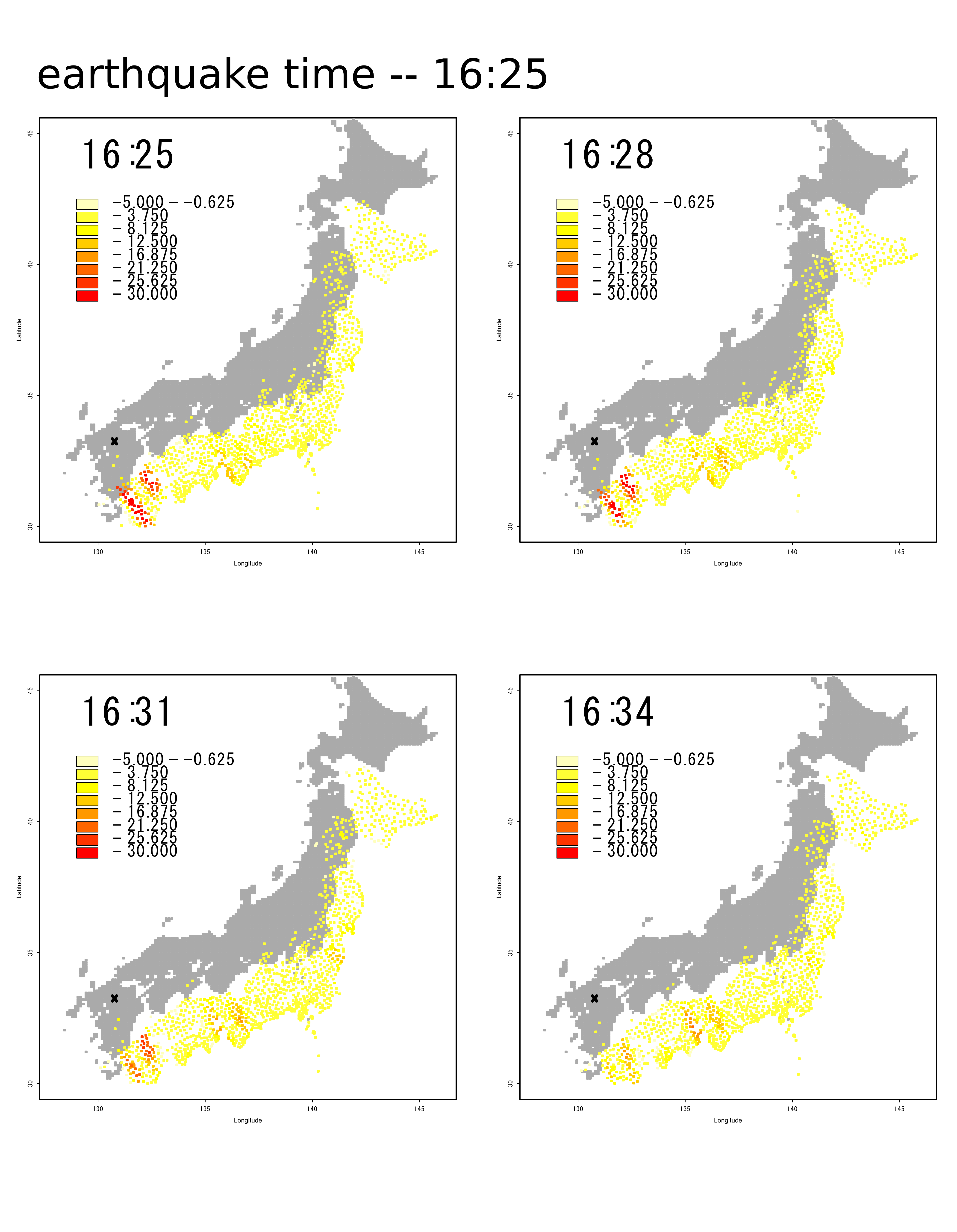}
\caption{Correlation values of PRN 17 over Japan after the main shock (16:25 UT).   The anomalous (red) area is getting vanished as time elapses.  The x mark represents the epicenter.}
\label{aftereq}
\end{figure}

Figure \ref{smallstep} shows the correlation values of PRN 17 over Japan with smaller time steps (3 minutes).
3 minutes time step is much smaller than the typical MSTID periods of 1000 seconds.
As seen in the figure, a part of anomalous area seems to propagate Southwestward, which is the typical MSTID behavior at nighttime and spring-summer seasons in several places of the world \cite{Hernandez2012}.
As a whole, however, the anomalous area stays near the epicenter.
This moveless behavior is unlike typical night-time MSTID.
Figure \ref{aftereq} shows the correlation values of PRN 17 after the earthquake.
The anomalous area near the epicenter is getting vanished as time elapses.

In the case of MSTIDs typically generated right after the earthquake, an ionospheric perturbation should propagates from the epicenter region as a circular MSTID \cite{Calais1995, Astafyeva2011, Cahyadi2015}.
The TEC anomalies before large earthquakes, however, do not propagate as a circular MSTID because the mechanism to be considered is radically different from the mechanism of MSTIDs after earthquakes \cite{Kuo2014,Pulinets2011}.

Figure \ref{mstid_20160101-02} shows the typical MSTIDs which are observed on 2016/01/01 and 2016/01/02.
During this period, the TEC anomalies can be seen sporadically in Japan and propagate southwestward as time elapses.
In the rest of this paper, we examine the differences of natures between MSTIDs (on 2016/01/01, 2016/01/02 and 2016/04/13) and TEC anomalies before the Kumamoto earthquake (2016/04/15).

\subsubsection{The anomalous area rates}
\begin{figure}
\includegraphics[width=0.8\linewidth]{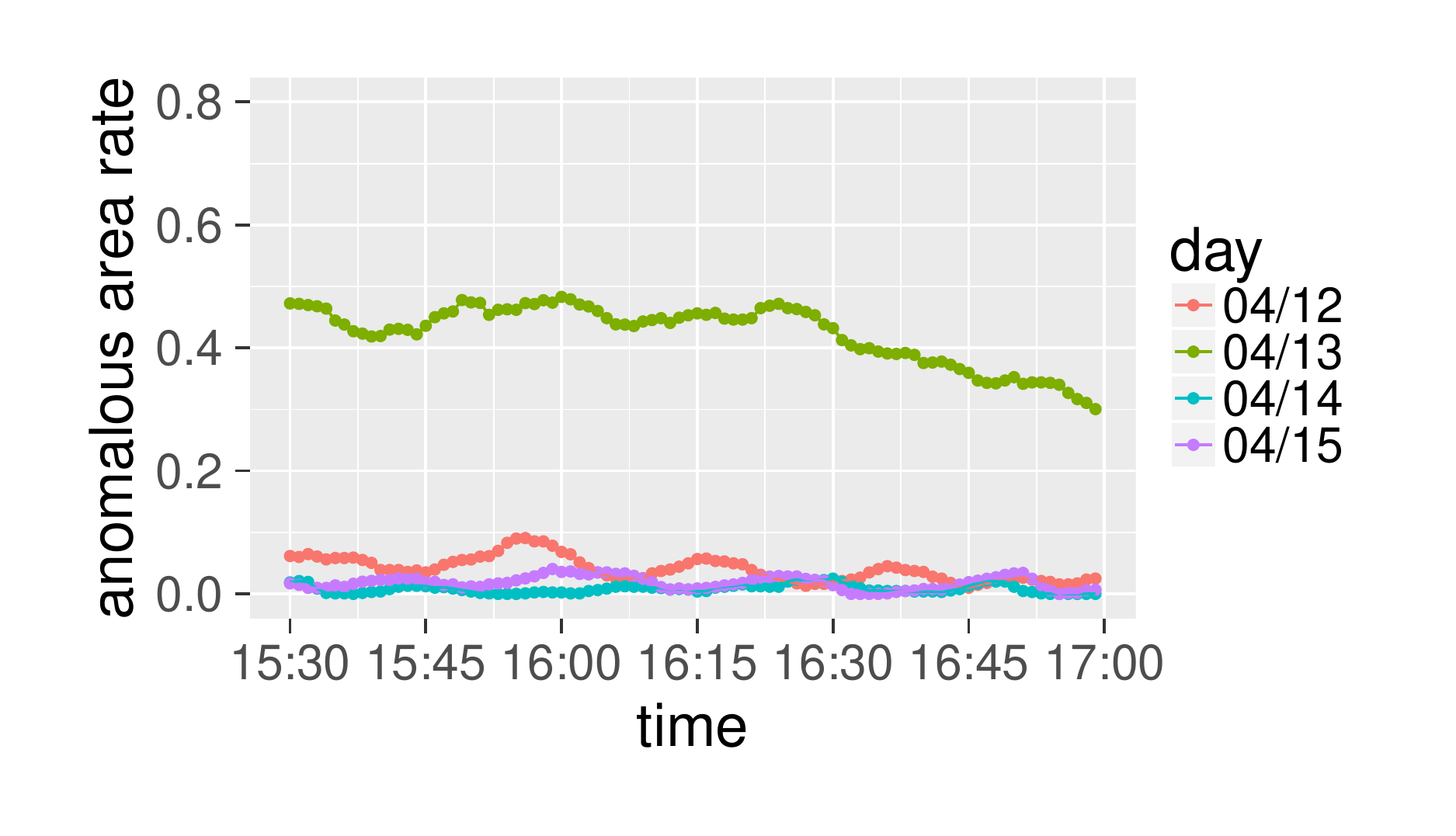}
\caption{The anomalous area rates from 2016/04/12 to 2016/04/15 in Japan.  The anomalous area rates mean the rates of GNSS stations whose correlation values $C(T)$ exceed the threshold, which is set up in advance.  The thereshold is 20 here.  We used the GPS satellite 17 in this figure. }
\label{anom_rate}
\end{figure}

As seen in Fig. \ref{non-earthquake} and Fig. \ref{mstid}, the ionospheric disturbences are captured on the non-earthquake day (2016/04/13).
We have to show the evidences which can prove clear differences between the TEC anomalies on the earthquake day (2016/04/15) and the non-earthquake day (2016/04/13).
In order to show that, we calculated the anomalous area rates in Japan at each time.
"The anomalous area rates" mean the rate of GNSS stations whose correlation values $C(T)$ exceed a certain threshold $\theta$, which is set up in advance.
The anomalous rate $r(T)$ is given by the following equation.
\begin{equation}
r(T) = num(\theta, T)/total\_num(T)
\end{equation}
Here, $num(\theta, T)$ is the number of GNSS stations whose $C(T)$ exceeds $\theta$ and $total\_num(T)$ is the total number of GNSS stations in Japan at time $T$.
In this paper, we set up $\theta = 20$.
Figure \ref{anom_rate} shows the comparison of $r(T)$ from 2016/04/12 to 2016/04/15.
It is clear that $r(T)$ on 2016/04/13 is very large in comparison to the other days, including the earthquake day.
This is because MSTIDs cause TEC anomalies in a relatively wide range, whereas the TEC anomalies before large earthquakes can occur in a relatively narrow range.

\begin{figure}
\includegraphics[width=0.75\linewidth]{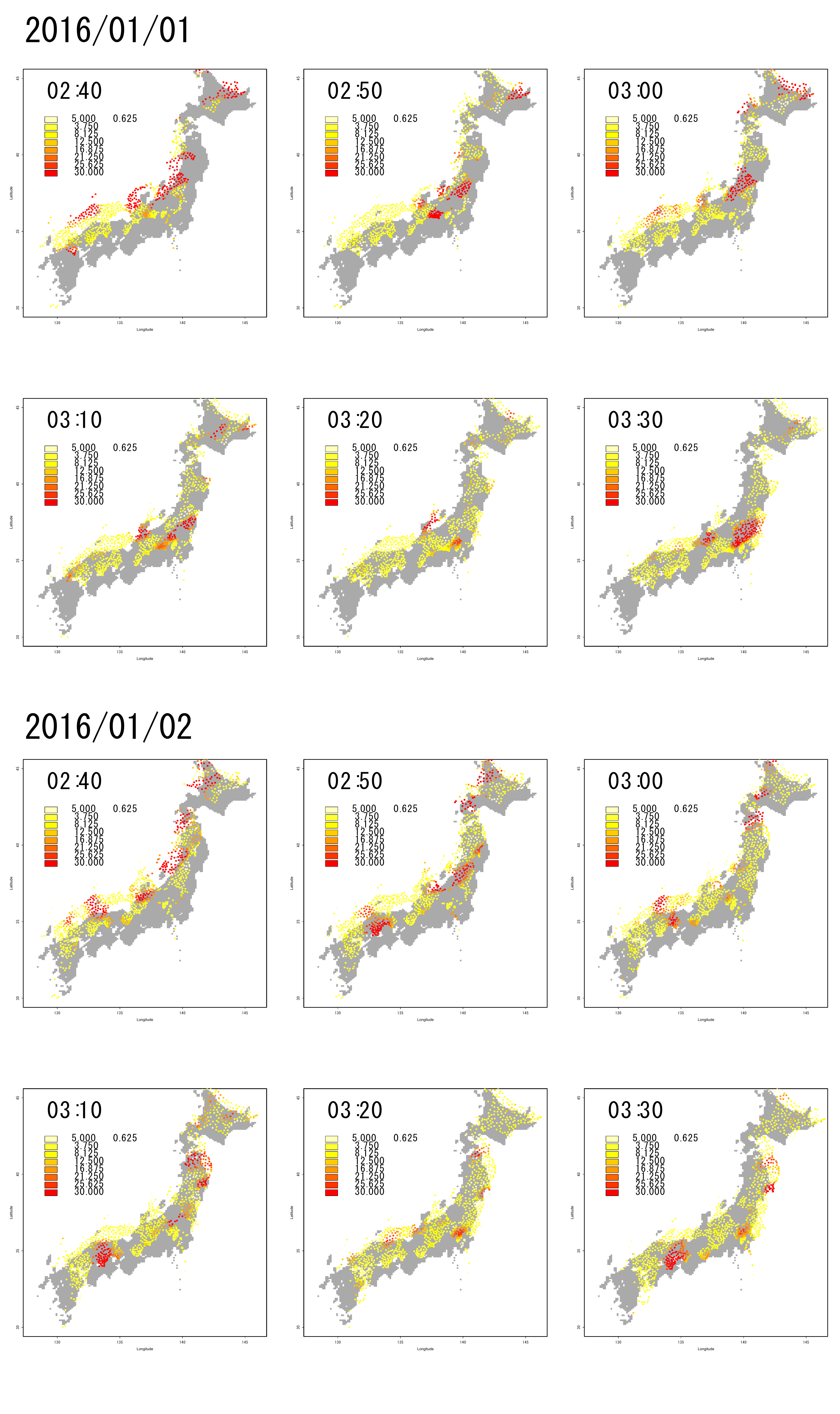}
\caption{Correlation values of PRN 5 over Japan on 2016/01/01 (above 2 rows) and 2016/01/02 (bottom 2 rows).  The time period is 02:40-03:30 in both cases.  Seasonal MSTIDs are captured by correlation analysis.  No large earthquakes occured in this period.}
\label{mstid_20160101-02}
\end{figure}

\begin{figure}
\includegraphics[width=0.8\linewidth]{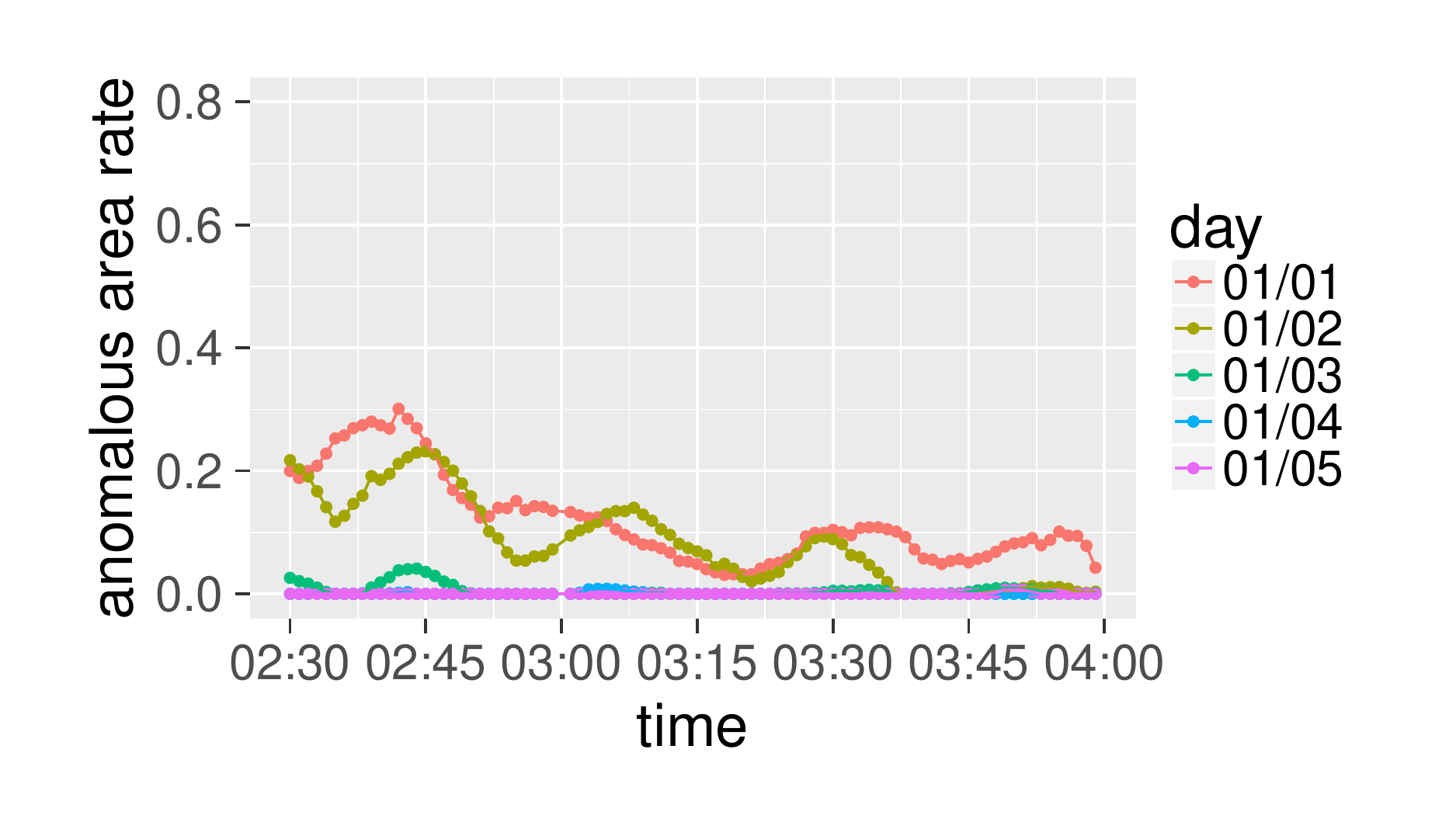}
\caption{The anomalous area rates from 2016/01/01 to 2016/01/05 in Japan.  The anomalous area rates mean the rates of GNSS stations whose correlation values $C(T)$ exceed the threshold, which is set up in advance.  The thereshold is 20 here.  We used the GPS satellite 5 in this figure. }
\label{anom_rate2}
\end{figure}

We also examined the anomalous area rates from 2016/01/01 to 2016/01/05.
2016/01/01 and 2016/01/02 are the days that MSTIDs are observed by correlation analysis as seen in Fig. \ref{mstid_20160101-02} and MSTIDs are not observed on the other days.
Figure \ref{anom_rate2} shows the comparison of $r(T)$ from 2016/01/01 to 2016/01/05.
$r(T)$ on 2016/01/01 and 2016/01/02 are slightly larger than those on the other days.

\subsubsection{The propagation velocities of the clustered TEC anomalies}
In order to show other differences of natures between MSTIDs and pre-seismic ionospheric anomalies, we investigated the propagation velocities of these TEC anomalies.
First of all, we classify the TEC anomalies into some clusters based on the geographical position.
Here, we defined the GNSS stations with $C(T) \geq 20$ as anomaly at each time $T$.
Next, we track and calculate the propagation velocities of the clustered TEC anomalies.
In this way, we investigated the propagation velocities of TEC anomalies on 2016/01/01 02:40-03:30, 2016/01/02 02:40-03:30, 2016/04/13 15:40-16:25 and 2016/04/15 15:40-16:25.

\begin{figure}
\includegraphics[width=0.8\linewidth]{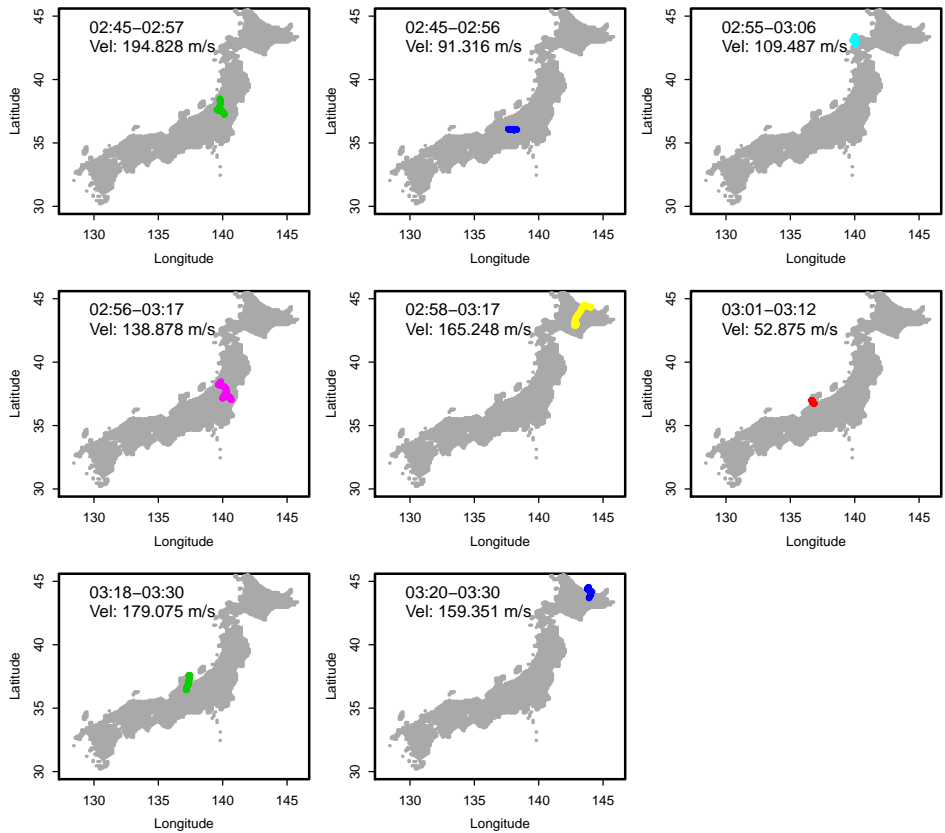}
\caption{The propagation velocities of the clustered TEC anomalies detected by correlation analysis on 2016/01/01 02:40 - 03:30.  The colored track in each window represents the track of clustered TEC anomaly's center.  The time period in which each clustered TEC anomaly is observed is described at the top left.  The propagation velocity of each clustered TEC anomaly is described at the underneath.}
\label{vel_20160101}
\end{figure}

Figure \ref{vel_20160101} shows the propagation velocities of TEC anomalies on 2016/01/01 02:40-03:30.
During this period, the TEC anomalies which are detected by correlation analysis are classified into 8 groups and each propagation velocity is calculated.
Though the TEC anomaly which is observed in 03:01-03:12 is relatively slow, other TEC anomalies show typical speeds of seasonal MSTIDs.

\begin{figure}
\includegraphics[width=0.8\linewidth]{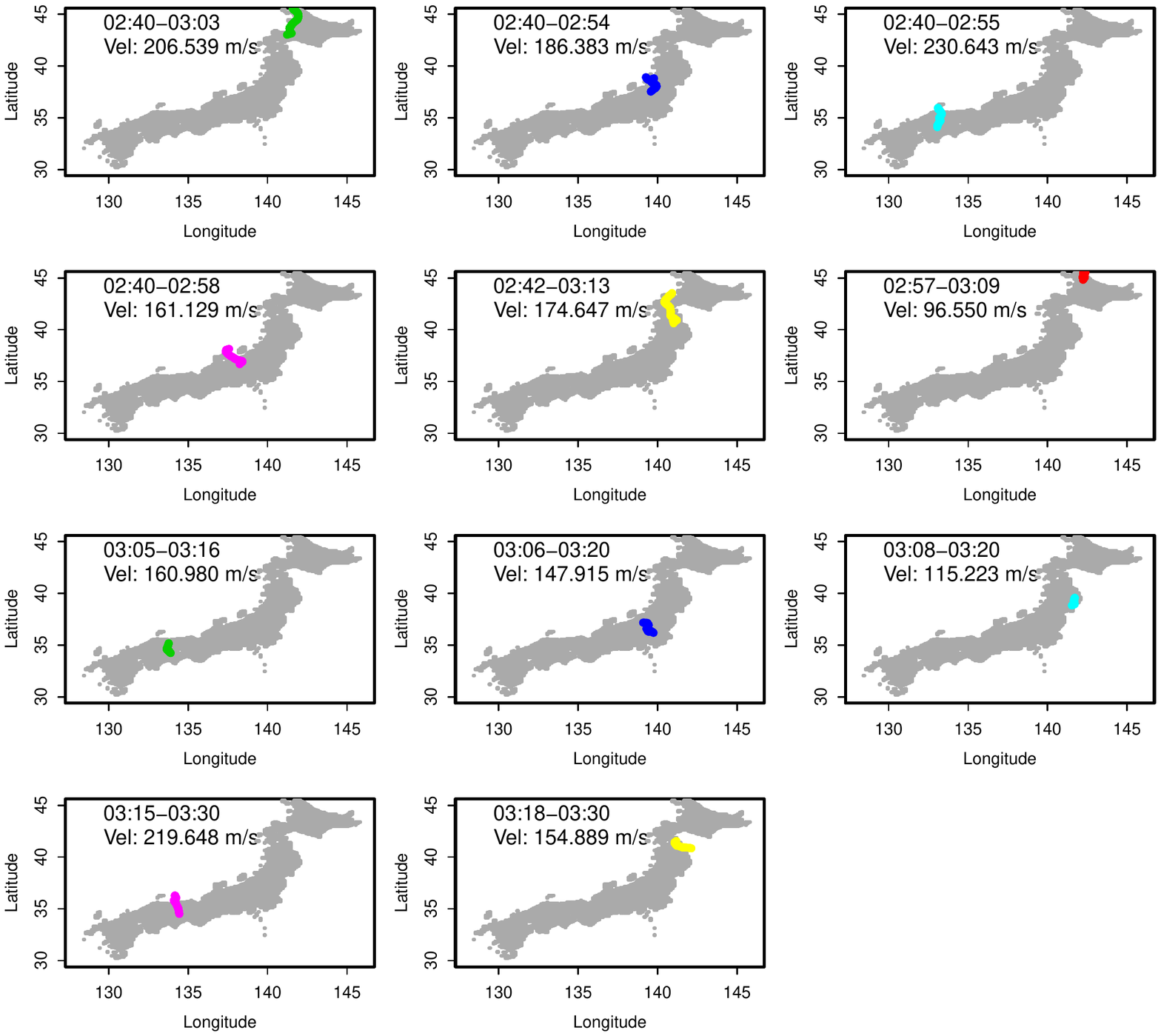}
\caption{The propagation velocities of the clustered TEC anomalies detected by correlation analysis on 2016/01/02 02:40 - 03:30.  The colored track in each window represents the track of clustered TEC anomaly's center.  The time period in which each clustered TEC anomaly is observed is described at the top left.  The propagation velocity of each clustered TEC anomaly is described at the underneath.}
\label{vel_20160102}
\end{figure}

Figure \ref{vel_20160102} shows the propagation velocities of TEC anomalies on 2016/01/02 02:40-03:30.
During this period, the TEC anomalies are classified into 11 groups and each propagation velocity is calculated.

\begin{figure}
\includegraphics[width=0.8\linewidth]{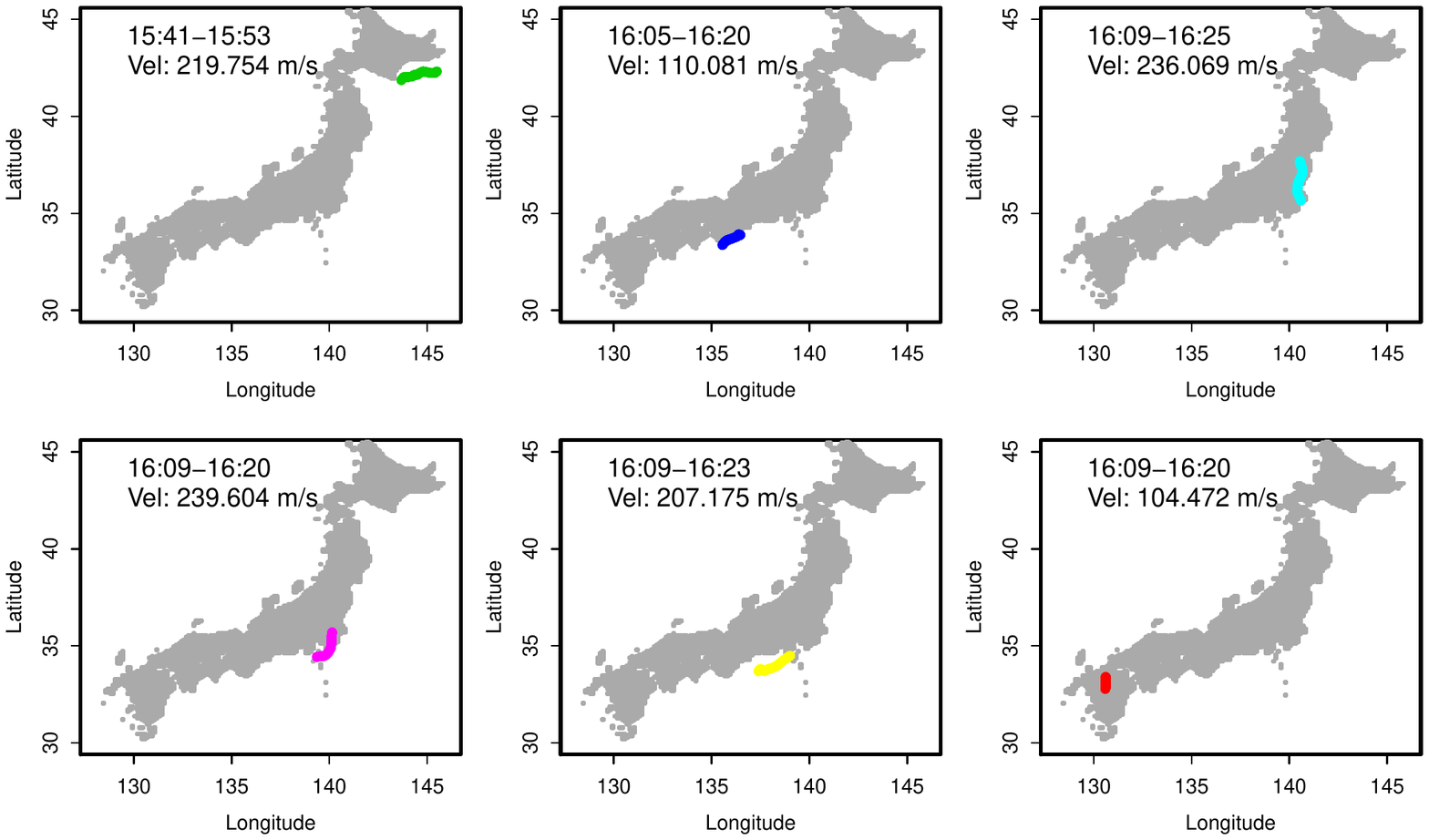}
\caption{The propagation velocities of the clustered TEC anomalies detected by correlation analysis on 2016/04/13 15:40 - 16:25.  The colored track in each window represents the track of clustered TEC anomaly's center.  The time period in which each clustered TEC anomaly is observed is described at the top left.  The propagation velocity of each clustered TEC anomaly is described at the underneath.}
\label{vel_20160413}
\end{figure}

Figure \ref{vel_20160413} shows the propagation velocities of TEC anomalies on 2016/04/13 15:40-16:25.
During this period, the TEC anomalies are classified into 6 groups and each propagation velocity is calculated.

\begin{figure}
\includegraphics[width=0.8\linewidth]{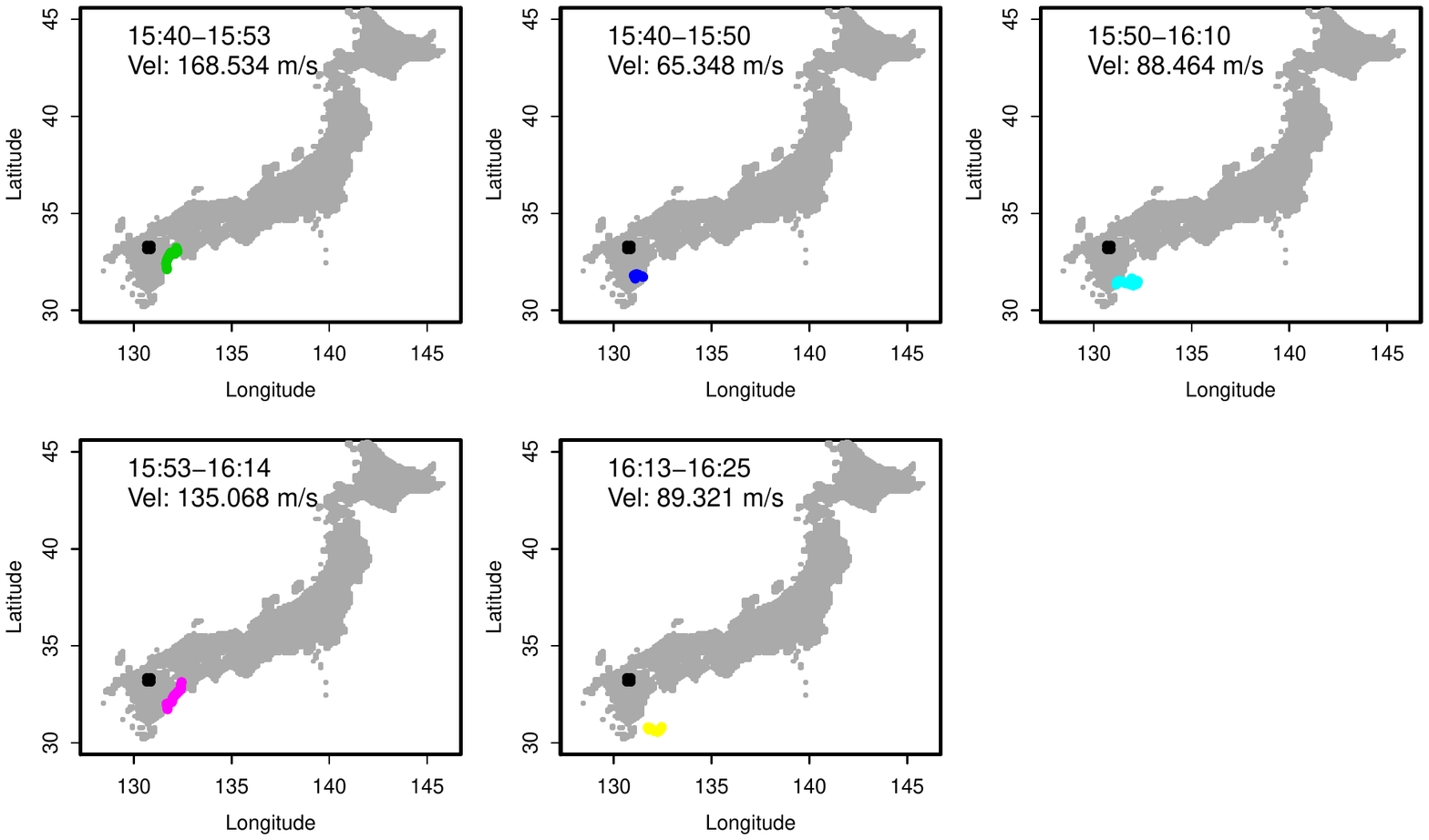}
\caption{The propagation velocities of the clustered TEC anomalies detected by correlation analysis on 2016/04/15 15:40 - 16:25.  The colored track in each window represents the track of clustered TEC anomaly's center.  The time period in which each clustered TEC anomaly is observed is described at the top left.  The propagation velocity of each clustered TEC anomaly is described at the underneath.}
\label{vel_20160415}
\end{figure}

\begin{figure}
\includegraphics[width=0.8\linewidth]{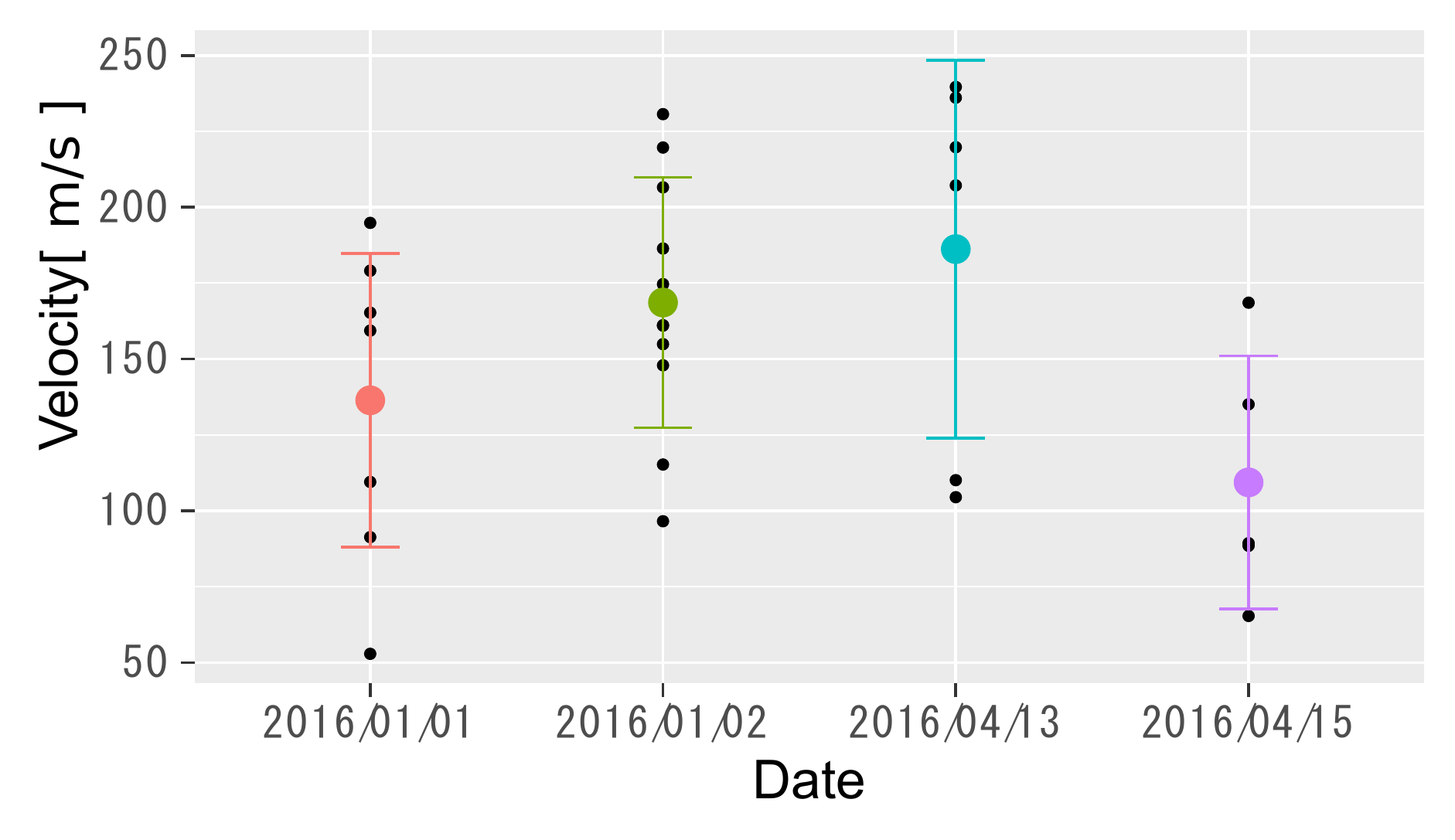}
\caption{The summary of Fig. \ref{vel_20160101} - \ref{vel_20160415}.  The propagation velocities in each figure are dotted in this figure.  The x-axis represents the date when the TEC anomalies are observed by correlation analysis.  The y-axis represents the velocities of the clustered TEC anomalies shown in Fig. \ref{vel_20160101} - \ref{vel_20160415}.  The colored dots and bar represent the mean values and the error bar, respectively.}
\label{scatter}
\end{figure}

Figure \ref{vel_20160415} shows the propagation velocities of TEC anomalies on 2016/04/15 15:40-16:25.
During this period, the TEC anomalies are classified into 5 groups and each propagation velocity is calculated.
We can confirm that some clustered TEC anomalies move slower than the typical seasonal MSTIDs observed on other days.
These slowly propagating TEC anomalies can be considered as pre-seismic TEC anomalies rather than seasonal MSTIDs.
Figure \ref{scatter} shows the summary of Fig. \ref{vel_20160101} - \ref{vel_20160415}.
As seen in this figure, the mean velocity on 2016/04/15 is lower than those on the other days.
In addition, more points which show the low velocity (less than 100 m/s) exist on 2016/04/15 (3 points of the 5 points).
These results suggest that some, if not all, TEC anomalies on 2016/04/15 behave differently from other MSTIDs in terms of propagation velocities.

As a further research, we need to determine whether the TEC anomalies detected before the Kumamoto earthquake are caused by (i) MSTID or (ii) compound of MSTID and the earthquake or (iii) the earthquake.
In order to detect the TEC anomalies before large earthquakes during the seasons in which MSTIDs appear frequently, the more detailed analysis supposing the case of the compound of MSTID and the preseismic TEC anomalies (case (ii)) is needed.
%------------------------------------------------

\section{Discussion}

In order to give a reasonable explanation to TEC anomalies before the 2016 Kumamoto earthquake, some more evidences in any on-shore earthquakes of Mw $\geq$ 7.0 are required (e.g. the 1995 Mw7.3 Kobe earthquake).
In this research, however, we could not get the GNSS data in such cases from GEONET because GEONET was not established but just planned in 1995.

In this research, we aimed only the TEC analysis method.
Our method showed TEC anomalies before the 2016 Kumamoto earthquake and this is one of the required condition to imply the causation between the pre-seismic activity and the TEC anomaly.
In order to examine the various aspects of the phenomena and lead more convincing conclusion, other geo-magnetic deflections need to be investigated in like manner with Heki and Enomoto \cite{Heki2013}.

The physical mechanism responsible for the TEC anomalies before large earthquakes is still unclear, but it has been researched so far and the physical models such as the coupling model for the lithosphere-atmosphere-ionosphere system which explain the preseismic TEC anomalies have been shown \cite{Kuo2011, Kuo2014, Pulinets2011}.
Kuo model assumed unusually large electric field, which was not observed in the 2016 Kumamoto earthquake, however.
Pulinets and Ouzounov model's approach is based on the most fundamental principles of tectonics giving understanding that earthquake is an ultimate result of relative movement of tectonic plates and blocks of different sizes.

The TEC anomalies detected by the correlation analysis before the earthquakes have some peculiar characteristics.
First, the TEC anomalies can be seen near the epicenter, but not just above the epicenter.
Second, as seen in Fig. \ref{correlation_local}, the abnormality time series shows a wave-like pattern.
This characteristic pattern also appears in the 2011 Tohoku-oki case \cite{Iwata2016}.
At this stage, there is no physical model which can explain these characteristics clearly.
Our research, however, revealed some remarkable traits (anomalous area rates and propagation velocities) of MSTIDs and pre-seismic TEC anomalies.
These data should be helpful for more understanding of the physical models for both of MSTIDs and pre-seismic ionospheric anomalies.
Furthermore, these characteristics are very interesting because they might be a clue to reveal the relation between pre-seismic ionosphere behavior and large earthquakes with Mw $\geq 7.0$ in the future.

%------------------------------------------------

\section{Conclusion}
In conclusion, the clear ionospheric anomalies are detected about one hour before the 2016 Kumamoto earthquake by the correlation analysis of TEC data.
As far as we know, this is the first pre-seismic observation of anomalies for the 2016 Kumamoto earthquake of Mw 7.3.
The pre-seismic TEC anomalies before the earthquake showed charasteristic patterns.
We also showed a method to distinguish the pre-seismic ionospheric disturbences and MSTIDs by considering the anomalous area rates.
Although we still do not know the physical mechanism which cause the pre-seismic ionospheric disturbences, these analysis results of TEC may help to understand the relationship between the ionosphere and earthquakes.
Further investigation of other large earthquakes and understanding of the physical mechanism of the pre-seismic ionospheric disturbances should be proceeded.


\begin{thebibliography}{}

\providecommand{\natexlab}[1]{#1}
\expandafter\ifx\csname urlstyle\endcsname\relax
  \providecommand{\doi}[1]{doi:\discretionary{}{}{}#1}\else
  \providecommand{\doi}{doi:\discretionary{}{}{}\begingroup
  \urlstyle{rm}\Url}\fi

\bibitem[{\textit{Astafyeva et al.}(2011)}]{Astafyeva2011}
Astafyeva, E. L., P. Lognonn\'{e}, and L. M. Rolland (2011), First ionospheric images of the seismic fault slip on the example of the Tohoku-Oki earthquake, \textit{Geophys. Res. Lett.}, \textit{38},  L22104.
\bibitem[{\textit{Cahyadi and Heki}(2015)}]{Cahyadi2015}
Cahyadi, M.N. and K. Heki (2015), Coseismic ionospheric disturbance of the large strike-slip earthquakes in North Sumatra in 2012: Mw dependence of the disturbance amplitudes, \textit{Geophys. J. Int.}, \textit{200}, 116--129.
\bibitem[{\textit{Calais and Minster}(1995)}]{Calais1995}
Calais, E., and J. B. Minster (1995), GPS detection of ionospheric perturbations following the January 17, 1994, Northridge earthquake, \textit{Geophys. Res. Lett.}, \textit{22}(9), 1045--1048.
\bibitem[{\textit{Dautermann et al.}(2009)}]{Dautermann2009}
Dautermann, T., E. Calais, and G. S. Mattioli (2009), Global Positioning System detection and energy estimation of the ionospheric wave caused by the 13 July 2003 explosion of the Soufrière Hills Volcano, Montserrat, \textit{Geophys. Res.},\textit{114}, B02202, \doi{10.1029/2008JB005722}.
\bibitem[{\textit{Donnelly}(1976)}]{Donnelly1976}
Donnelly, R. F. (1976), Empirical Models of Solar Flare X Ray and EUV Emission for Use in Studying Their E and F Region Effects, \textit{J. Geophys. Res.}, \textit{81}(25),4745--4753.
\bibitem[{\textit{Heki}(2006)}]{Heki2006}
Heki, K. (2006), Explosion energy of the 2004 eruption of the Asama Volcano, Central Japan, inferred from ionospheric disturbance, \textit{Geophys. Res. Lett.}, \textit{33}, L14303.
\bibitem[{\textit{Heki}(2011)}]{Heki2011}
Heki, K. (2011), Ionospheric electron enhancement preceding the 2011 Tohoku-Oki earthquake, \textit{Geophys. Res. Lett.}, \textit{38}.  L17312.
\bibitem[{\textit{Heki and Enomoto}(2013)}]{Heki2013}
Heki, K. and Y. Enomoto (2013), Preseismic ionospheric electron enhancements revisited, \textit{J. Geophys. Res. Space Phys.}, \textit{118}, 6618--6626.
\bibitem[{\textit{Heki and Enomoto}(2015)}]{Heki2015}
Heki, K. and Y. Enomoto (2015), Mw dependence of preseismic ionospheric electron enhancements, \textit{J. Geophys. Res. Space Phys.}, \textit{120}, 7006--7020, \doi{10.1002/2015JA021353}.
\bibitem[{\textit{Ho et al.}(1996)}]{Ho1996}
Ho, C. M., A. J. Mannucci, U. J. Lindqwister, X. Pi, and B. T. Tsurutani (1996), Global ionosphere perturbations monitored by the worldwide GPS network, \textit{Geophys. Res. Lett.}, \textit{23}, 3219-3222.
\bibitem[{\textit{Hern\'{a}ndez-Pajares et al.}(2012)}]{Hernandez2012}
Hern\'{a}ndez-Pajares, M., Juan, J. M., Sanz, J., and Arag\'{o}n-\`{A}ngel, A. (2012), Propagation of medium scale traveling ionospheric disturbances at different latitudes and solar cycle conditions, \textit{Radio Science}, \textit{47}(6), \doi{10.1029/2011RS004951}.
\bibitem[{\textit{Iwata and Umeno}(2016)}]{Iwata2016}
Iwata, T and K. Umeno (2016), Correlation Analysis for Pre-seismic Total Electron Content Anomalies around the 2011 Tohoku-Oki Earthquake, \textit{J. Geophys. Res. Space Phys.}, \textit{121}, \doi{10.1002/2016JA023036}.
\bibitem[{\textit{Kamogawa and Kakinami}(2011)}]{Kamogawa2011}
Kamogawa, M. and Y. Kakinami. (2011), Is an ionospheric electron enhancement preceding the 2011 Tohoku-oki earthquake a precursor?, \textit{J. Geophys. Res. Space Phys.}, \textit{118}(4), 1751--1754.
\bibitem[{\textit{Kuo et al.}(2011)}]{Kuo2011}
Kuo, C. L., J. D. Huba, G. Joyce, and L. C. Lee (2011), Ionosphere plasma bubbles and density variations induced by pre-earthquake rock currents and associated surface charges, \textit{J. Geophys. Res.}, \textit{116}, A10317, \doi{10.1029/2011ja016628}.
\bibitem[{\textit{Kuo et al.}(2014)}]{Kuo2014}
Kuo, C. L., L. C. Lee, and J. D. Huba (2014), An improved coupling model for the lithosphere-atmosphere-ionosphere system, \textit{J. Geophys. Res. Space Physics}, \textit{119}, 3189--3205, \doi{10.1002/2013JA019392}.
\bibitem[{\textit{Leonovich}(2002)}]{Leonovich2002}
Leonovich, L. A., E. L. Afraimovich, E. B. Romanova, and A. V. Taschilin (2002), Estimating the contribution from different ionospheric regions to the TEC response to the solar flares using data from the international GPS network, \textit{An. Geophys.}, \textit{20}, 1935-1941.
\bibitem[{\textit{Masci et al.}(2015)}]{Masci2015}
Masci, F., J. N. Thomas, F. Villani, J. A. Secan, and N. Rivera (2015), On the onset of ionospheric precursors 40 min before strong earthquakes, \textit{J. Geophys. Res. Space Phys.}, \textit{120}(2), 1383--1393.
\bibitem[{\textit{Mannucci et al.}(1998)}]{Mannucci1998}
Mannucci, A. J., B. D. Wilson, D. N. Yuan, C. H. Ho, U. J. Lindqwister, and T. F. Runge (1998), A global mapping technique for GPS-derived ionospheric total electron content measurements, \textit{Radio Sci.}, \textit{33}, 565-582.
\bibitem[{\textit{Otsuka et al.}(2011)}]{Otsuka2011}
Otsuka, Y., Kotake, N., Shiokawa, K., Ogawa, T., Tsugawa, T., and Saito, A (2011), Statistical study of Medium-Scale Traveling Ionospheric Disturbances observed with a GPS receiver network in Japan, \textit{IAGA Special Sopron Book Series}, Vol. 2, Part 3, 291--299, \doi{10.1007/978-94-007-0326-1 21}.
\bibitem[{\textit{Pulinets and Ouzounov}(2011)}]{Pulinets2011}
Pulinets, S., and Ouzounov, D. (2011), Lithosphere-Atmosphere-Ionosphere Coupling (LAIC) model-An unified concept for earthquake precursors validation, \textit{Journal of Asian Earth Sciences}, \textit{41} (4), 371-382.
\bibitem[{\textit{Saito et al.}(1998)}]{Saito1998}
Saito, A., S. Fukao. and S. Miyazaki (1998), High resolution mapping of TEC perturbations with the GSI GPS network over Japan, \textit{Geophys. Res. Lett.}, \textit{25}, 3079-3082.
\bibitem[{\textit{Sard\'{o}n}(1997)}]{Sardon1997}
Sard\'{o}n, E. and N. Zarraoa (1997), Estimation of total electron content using GPS data: How stable are the differential satellite and receiver instrumental biases?, \textit{Radio Sci.}, \textit{32}, 1899-1910.
\bibitem[{\textit{Utada and Shimizu}(2011)}]{Utada2011}
Utada, H., and H. Shimizu (2011), Comment on ‘Preseismic ionospheric electron enhancements revisited’ by K. Heki and Y. Enomoto, \textit{J. Geophys. Res. Space Phys.}, \textit{119}(7), 6011--6015.

\end{thebibliography}
\end{document}